\documentclass[journal,transmag]{IEEEtran}
\usepackage{ifpdf}
\usepackage{cite}
\usepackage{graphicx}
\usepackage{epsfig}
\usepackage[dvips]{color}

\usepackage[cmex10]{amsmath}
\usepackage{array}

\begin{document}
\title{Polarizability Tensor Calculation:\\Induced Local Charge and Current Distributions}

\author{\IEEEauthorblockN{Mohammad Yazdi\IEEEauthorrefmark{1},
Mohammad Albooyeh\IEEEauthorrefmark{2},~\IEEEmembership{Member,~IEEE,}
Nader Komjani\IEEEauthorrefmark{1}, and
Constantin Simovski\IEEEauthorrefmark{2}}
\IEEEauthorblockA{\IEEEauthorrefmark{1}Department of Electrical Engineering, Iran University of Science and Technology, Narmak, Tehran, Iran}
\IEEEauthorblockA{\IEEEauthorrefmark{2}Department of Radio Science and Engineering, SMARAD Centre of Excellence, Aalto University, Aalto, Finland}
}

\IEEEtitleabstractindextext{%
\begin{abstract}
We develop a semi-analytical approach to calculate the polarizability tensors of an arbitrary individual scatterer. The approach is based on the calculation of induced electric and/or magnetic dipole moments on the scatterer. By taking the advantages of the present approach, we calculate the individual polarizability tensors of an isolated scatterer in a homogeneous isotropic medium. Moreover, we obtain the polarizability tensors of scatterers located between two different isotropic media. Furthermore, due to the nature of the proposed method, we may determine the effective polarizability of a scatterer in a periodic array. To this end and for comparison reasons, we investigate two other methods for calculating effective polarizability tensors of scatterers in a two-dimensional periodic array. The proposed approach, in comparison to other reported approaches, is simpler, easily implemented, and does not require spherical harmonic expansion or complicated far-field calculations. We examine the validity of the proposed approach using several examples and compare the results with other approaches.
\end{abstract}

}

\maketitle

\IEEEdisplaynontitleabstractindextext

\IEEEpeerreviewmaketitle

\section{Introduction}
\IEEEPARstart{T}{he} electromagnetic response of composite materials, including metamaterials, mainly depends on two important factors: 1) the properties of their ``constituent elements''~\footnote{Also, called ``inclusion'', ``scatterer'', or ``particle'' in the current manuscript. It is called ``meta-atom'' in metamaterials.} and 2) the manner in which these elements are arranged together. The latter determines the interaction effects between the elements while the former demonstrates the general response of the composite. Indeed, the electromagnetic behaviour of the constituent elements may be described by the polarizability tensors~\cite{Serdyukov}. Polarizability tensors determine the relation between the induced dipole moments and the external electromagnetic field excitation for a particle with linear response to the fields~\cite{Tretyakov1}. We may sufficiently explain the electromagnetic behaviour of many small particles by their dipolar moments or equivalently their polarizability tensors. Therefore, various \textit{analytical} and \textit{numerical} approaches have been employed to calculate the polarizability tensors of different particles ~\cite{Sergei_Chiral, Mirmoosa, Rasoul, Arango, Ishimaru_Generalized}.

For instance, the polarizability of a dielectric sphere is analytically calculated in~\cite{Ishimaru_Scattering}. Moreover, analytic formulas are derived for polarizabilities of wire dipole~\cite{Tretyakov1}, chiral~\cite{Sergei_Chiral}, and $\Omega$-shaped particles~\cite{Simovski_Omega}. Furthermore, the polarizabilities of two nonreciprocal magnetic particles; i.e., Tellegen-omega and ``moving''-chiral are formulated in~\cite{Mirmoosa}. The proposed analytical approaches are not only difficult but also limited to simple structures. Moreover, small modifications to the particle structure {result} in a different analytical model.

Consequently, for more complicated shapes, different semi-analytical and numerical methods have been proposed~\cite{Rasoul, Arango, Terekhov, Ishimaru_Generalized}. For example, Alaee \textit{et al.} derived the polarizabilties of a planar omega structure using multipole moments approach~\cite{Rasoul}. Their method was based on the expansion of spherical harmonics of the Mie coefficients~\cite{Bohren} in the Cartesian coordinate system. 
One of the disadvantages of this method is the cumbersome calculations of the spherical harmonics. Furthermore, the method is only valid for particles in homogeneous host medium~\cite{Mulig}.

Recently, another numerical method is developed to retrieve the polarizability tensors of general bianisotropic particles~\cite{Viktar}. It directly calculates the induced dipole moments of an excited scatterer and does not require complicated calculations and harmonic expansions. In this method, one probes the scattered far-fields of an excited particle and relates them to the radiation effects of a pair of electric and magnetic dipole moments. This method is based on the assumption of predominantly dipolar resonances and neglecting higher order multipoles. Although this method is comprehensive, it is not suitable for particles in inhomogeneous host media (such as substrated particles). Moreover, with this method, it is impossible to directly calculate the \textit{effective} polarizability tensors of particles in periodic arrays. In a periodic array, the effective polarizability tensors are parameters which take the effect of the interaction of particles into account. Indeed, using the proposed method in~\cite{Viktar} one needs to separately calculate the \textit{interaction} between the particles and then analytically connect them to the calculated \textit{individual} polarizability tensors in order to determine the effective polarizability tensors.

To overcome all the above deficiencies, we propose a method to calculate the polarizability tensors of a general bianisotropic particle. This method is based on the calculation of the near field response of a particle to the plane wave illumination. We use the induced local charges/currents on the particle to determine multipole moments and consequently the polarizability tensors. This way we provide an approach in which it does not only add the strength of different reported semi-analytical approaches in~\cite{Rasoul, Viktar}\footnote{That is, calculation of higher order multipoles as well as the polarizability retrieval of general bianisotropic particles.} but also enables one to determine the polarizabilities of particles located on substrates. Furthermore, with our approach we demonstrate how one may consider the case in which a particle is surrounded by similar particles. That is, how one can directly calculate the effective polarizability tensors of particles in planar periodic arrays.

Hence, the paper is organized as following. The second section introduces our approach: the general ideas, formulation and numerical test set-up. It provides the procedure of evaluating polarizability tensors for an arbitrary isolated particle in a homogenous host medium such as free space. It is followed up by the considerations needed for substarted particles. Then, our approach is investigated to calculate the effective polarizabilities of particles in planar periodic arrays. Next, the concept of rotation matrix is introduced. At the end of this section, two more distinct methods for calculating the effective polarizabilities in arrays are presented for the sake of comparison. After presenting the analytical framework, in the third section, the proposed method is verified using various examples.

In the third section, we start with the calculation of polarizability tensors of an individual chiral particle in free space and then compare the results with the analytical method presented in~\cite{Sergei_Chiral} as well as the scattered far-field method presented in~\cite{Viktar}. Then, we calculate the so called effective polarizabilities of these chiral particles when they are periodically positioned in a planar surface. Notice, the calculation of effective polarizabilities is impossible using the presented approaches in~\cite{Rasoul, Viktar} since they use the scattered far-fields as their input data. We also compare the results with two other proposed methods introduced in section~\ref{theory_section_2}. Next, we examine a non-reciprocal Tellegen-omega particle and compare its polarizabilities with those presented in~\cite{Viktar}. Thereafter, we calculate the polarizability tensors of a nano-gold split ring resonator to demonstrate the capability of our proposed method in the terahertz frequency region. We also show that the higher order multipoles of such structure are negligible compared to its dipole moments. This is impossible to demonstrate with the presented method in~\cite{Viktar}. As the last example, we calculate the polarizabilitis of a truncated sphere on a substrate. This is impossible to calculate using the proposed semi-analytical methods in~\cite{Rasoul, Viktar}. The results are compared with the analytical methods presented in ~\cite{Wind, Pederson}. We show that the presented analytical methods can only provide the electric polarizabilities of this structure and do not present the bianisotropic terms; whereas it is proven in ~\cite{SIB, Albooye_SIB} that the presence of the substrate may induce bianisotropy to the structure. Therefore, one needs to consider the bianisotropic terms in the calculation of polarizabilities as we perform in our approach. Finally, in the last section we conclude the study.

\section{Theory}
We first present a procedure for retrieval of the \textit{individual} polarizability tensors of an arbitrary scatterer in a homogeneous host medium. Then, by introducing the concept of rotation matrix, we extend the proposed procedure in order to be suitable for the substrated scatterers. Moreover, we expand our method in order to calculate the {\textit{effective}} polarizability tensors of scatterers in {planar periodic arrays}. At the end of this section and aside the proposed method, we develop two different methods to calculate the {\textit{effective}} polarizabilities of scatterers in planar periodic arrays.

\subsection{Extraction of the Polarizability Tensors}
The parameters which connect the induced electric and magnetic dipole moments $\textbf{p}$ and $\textbf{m}$ to the ``local'' electric and magnetic fields $\textbf{E}^\textrm{loc}$ and $\textbf{H}^\textrm{loc}$ are called polarizabilities. These relations may be summarized in the following compact forms~\cite{Tretyakov1}:\begin{equation}
\begin{bmatrix}
      \textbf{p}\\
      \textbf{m}\\
    \end{bmatrix}=\begin{bmatrix}
      \bar{\bar{\alpha}}^\textrm{ee} & \bar{\bar{\alpha}}^\textrm{em}\\
      \bar{\bar{\alpha}}^\textrm{me} & \bar{\bar{\alpha}}^\textrm{mm}\\
    \end{bmatrix}\begin{bmatrix}
      \textbf{E}^\textrm{loc}\\
      \textbf{H}^\textrm{loc}\\
    \end{bmatrix}.
\label{defpm}
\end{equation}Here $\bar{\bar{\alpha}}^\textrm{ee}$, $\bar{\bar{\alpha}}^\textrm{mm}$, $\bar{\bar{\alpha}}^\textrm{em}$, and $\bar{\bar{\alpha}}^\textrm{me}$ are the electric, magnetic, magnetoelectric, and electromagnetic polarizability tensors, respectively. In a three-dimensional coordinate system, each of these polarizability tensors has nine components. Therefore, in order to completely characterize a particle response to an electromagnetic wave, there are $36$ polarizability components that should be calculated.

\begin{figure*}
\centering
\includegraphics[scale=0.25]{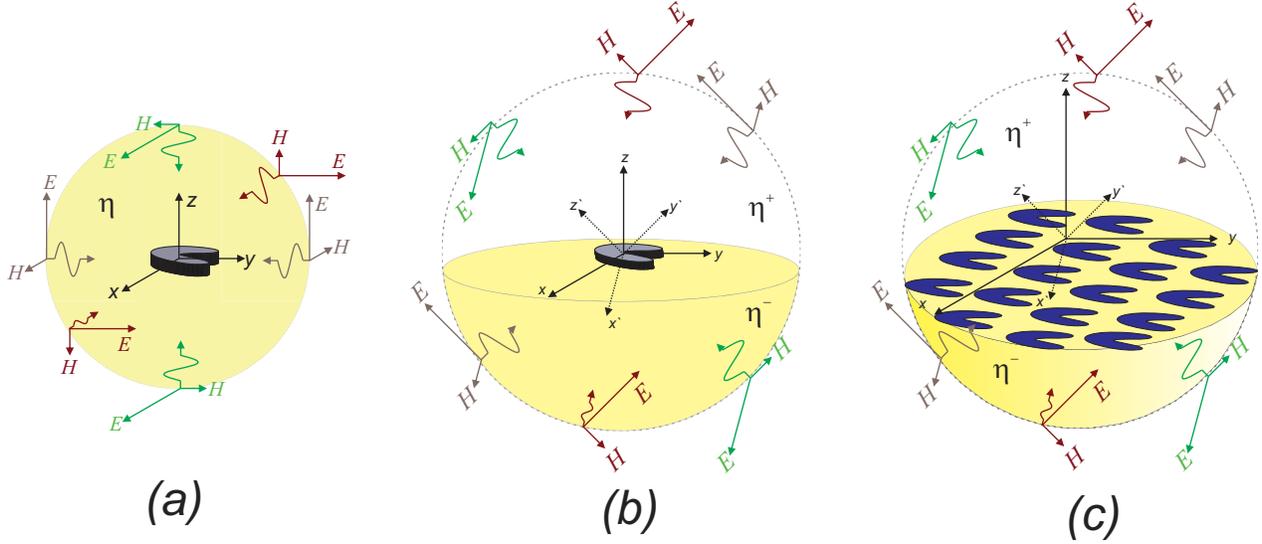}
\caption{Polarizability tensor extraction setup: a) an arbitrary scatterer is located in the origin of the coordinates and is illuminated by six orthogonal plane waves, b) an arbitrary scatter is located on a half space substrate, the six plane waves are rotated so that can excite the substrated particle, c) an array in the x-y plane that uses same setup as in (b).}
\label{test_setup}
\end{figure*}

Let us first consider a particle in the origin of the Cartesian coordinate system in a homogeneous host medium with the intrinsic impedance $\eta$ as shown in Fig.~\ref{test_setup}-a. Now let us start with the calculation of $\alpha^\textrm{ee}_\textrm{ix}$, and  $\alpha^\textrm{me}_\textrm{ix}$ where $\textrm{i=x,y,z}$. According to Ref. \cite{Terekhov} and equation (\ref{defpm}), we should determine the particle response ($\textbf{p}_i$ and $\textbf{m}_i$) to an $x$-directed exciting electric field. To create such an electric field, we apply two oppositely plane waves which are traveling in the $z$-direction with similar polarizations for their electric fields (green color vectors in Fig.~\ref{test_setup}-a); i.e., $E_x^\pm=E_0 e^{j(\omega t\mp k z)}$, where $\omega$ and $k$ are the angular frequency and wavenumber, respectively~\footnote{we consider the time dependence $e^{j\omega t}$ all over the manuscript}. Notice, $\pm$ sign in the superscript refers to the propagation in $\pm z$-direction. Since the two magnetic fields are out of phase, then summing these two plane waves copies a standing wave with an electric-field maximum and a magnetic-field zero at the place of the particle assuming the particle is small compared to the wavelength. As a result, the polarizabilities can be easily calculated as:
\begin{equation}
\begin{matrix}
  \alpha_\textrm{ix}^\textrm{ee}=\frac{ \textrm{p}_\textrm{i}^+ + \textrm{p}_\textrm{i}^-}{2 E_0} ,&&&&\alpha_\textrm{ix}^\textrm{me}=\frac{ \textrm{m}_\textrm{i}^+ +  \textrm{m}_\textrm{i}^-}{2 E_0} \\
   \end{matrix}.
     \label{alpha1}
\end{equation}Here, the $\pm$ sign in the superscript of the dipole moments, denotes for the induced moment by the corresponding plane wave in $\pm z$-direction. In a similar way, subtracting these two plane waves creates a local maximum of the magnetic field with zero electric field. Therefore, one can write
\begin{equation}
\displaystyle
\begin{matrix}
  \alpha_\textrm{iy}^\textrm{em}=\frac{\textrm{p}_\textrm{i}^+ - \textrm{p}_\textrm{i}^-}{2 E_0}{\eta}, &&\alpha_\textrm{iy}^\textrm{mm}=\frac{\textrm{m}_\textrm{i}^+ - \textrm{m}_\textrm{i}^-}{2 E_0}\eta \\
   \end{matrix}.
        \label{alpha2}
\end{equation}

We should mention again that the exciting field is assumed to be a constant value in the place of the particle. This is true when the size of the particle is sufficiently smaller than the operating wavelength.

One may obtain 12 polarizabilty components from equations (\ref{alpha1}) and (\ref{alpha2}). Other components may be calculated similarly with $E_y^\pm=E_0 e^{j(\omega t\mp kx)}$ and $E_z^\pm=E_0 e^{j(\omega t\mp ky)}$ as the excitation electromagnetic wave (red color and brown color vectors in Fig.~\ref{test_setup}-a).
So far, we expressed that six orthogonal plane waves are needed to calculate all polarizability components. Moreover, we built the relations between polarizability components and the dipole moments. In the next step, we present how the required dipole moments may be calculated from the induced currents and/or charges~\cite{Raab}.

When an electromagnetic field illuminates a particle, the equilibrium of charges (either free charges or  bounded ones) in/on the particle is changed. This excited particle can be modelled as an expansion of multipole moments. The first four successive electric multipole moments in terms of the induced charge density $\rho (\textbf{r})$ are expressed as~\cite{Raab}\begin{equation}
\begin{array}{ll}
\begin{matrix}
\textrm{q} =  \int_V \rho (\textbf{r}) dv,&&&&\textrm{p}_\textrm{i} =  \int_V r_\textrm{i} \rho (\textbf{r}) dv,\\
\\
\textrm{q}_\textrm{ij} =  \int_V r_\textrm{i} r_\textrm{j} \rho (\textbf{r}) dv.\\
\end{matrix}
\end{array}
\label{defem}
\end{equation}
{Where $V$ is the volume of the particle and $r$ is the position vector in the Cartesian coordinate system. Subscripts} i,j,k denote components of a Cartesian tensor, and $\textrm{q}$, $\textrm{p}_\textrm{i}$, and $\textrm{q}_\textrm{ij}$ are the electric monopole, dipole, and quadrupole moments, and so forth, respectively. Since the total electric charge of the particle after the illumination remains zero, we have: $q = 0$.
\\Similarly, the first two magnetic multipole moments are expressed in terms of the induced current density $\textbf{J} (\textbf{r})$ as:\begin{equation}
\begin{array}{ll}
\textrm{m}_\textrm{i} = \frac{1}{2} \int_V (\textbf{r} \times \textbf{J} (\textbf{r}))_\textrm{i}  dv,\\
\\
\textrm{m}_\textrm{ij} = \frac{2}{3} \int_V (\textbf{r} \times \textbf{J} (\textbf{r}))_\textrm{i} r_\textrm{j}  dv,\\
\\\end{array}
\label{defmm}
\end{equation}where $\textrm{m}_\textrm{i}$, and $\textrm{m}_\textrm{ij}$ are the magnetic dipole and quarupole moments, respectively.
Finally, we may summarize our procedure as following. First, we separately subject the scatterer to three couples of plane waves in three orthogonal directions. Then, using the dipole moments calculated from equations (\ref{defem}) and (\ref{defmm}) we may easily find the polarizability components from equations (\ref{alpha1}) and (\ref{alpha2}).

\subsection{Rotation Matrix}
\label{rot_mat}
With the proposed method we require the wave propagation in three orthogonal directions to determine the whole polarizability tensors of a particle. Therefore, this approach can be easily applied to any arbitrary scatterer in a homogeneous host medium. However, in the case of substrated scatterers (Fig.~\ref{test_setup}-b) or scatterers in planar arrays (Fig.~\ref{test_setup}-c), one cannot attribute a specific wave number to the waves propagating along the interface of the substrate or the array. Therefore, the interface plane is forbidden in our methodology and one may not simply use equations (\ref{alpha1}) and (\ref{alpha2}) for the waves propagating in the directions along the interface; that is, $x$ and $y$ in Figs.~\ref{test_setup}-b,~\ref{test_setup}-c. However, since the polarizabilities of particles are characteristic parameters of particles, then they should not depend on the propagation direction. As a result, we propose to rotate the excitation fields so that there is no incident wave along the interface.

Now, let us first assume the particle or the array is located at the interface of two different media with the intrinsic impedances $\eta^+$ and $\eta^-$ as shown in Figs.~\ref{test_setup}-b or~\ref{test_setup}-c. The interface is assumed to be at the $xy$-plane. We then rotate the coordinate of the incident waves and find the polarizability tensors in the rotated coordinate system ($x'-y'-z'$). After that, we perform the same procedure for two traveling waves in the opposite directions $\pm z'$; i.e., $E_x'^\pm=E_0 e^{j(\omega t\mp k^\mp z')}$ (green vectors in Fig.~\ref{test_setup}-b). This way we obtain the polarizability components as we did in equations (\ref{alpha1}) and (\ref{alpha2}), but now in the rotated coordinate system; i.e., $\alpha^\textrm{ee}_{\textrm{ix}'}$, $\alpha^\textrm{me}_{\textrm{ix}'}$, $\alpha_{\textrm{iy}'}^\textrm{em}$, and $\alpha_{\textrm{iy}'}^\textrm{mm}$ with $\textrm{i=} \textrm{x}', \textrm{y}', \textrm{z}'$:\begin{equation}
\begin{matrix}
  \alpha_{\textrm{ix}'}^\textrm{ee}=\frac{\eta^- \textrm{p}_\textrm{i}^+ +\eta^+- \textrm{p}_\textrm{i}^-}{(\eta^- + \eta^+) E_0},&\alpha_{\textrm{ix}'}^\textrm{me}=\frac{\eta^- \textrm{m}_\textrm{i}^+ + \eta^+ \textrm{m}_\textrm{i}^-}{(\eta^- + \eta^+)E_0}, \\
  \\
   \alpha_{\textrm{iy}'}^\textrm{em}=\frac{\textrm{p}_\textrm{i}^+ -\textrm{p}_\textrm{i}^-}{(\eta^- + \eta^+) E_0}{\eta^- \eta^+}, &\alpha_{\textrm{iy}'}^\textrm{mm}=\frac{\textrm{m}_\textrm{i}^+ - \textrm{m}_\textrm{i}^-}{(\eta^- + \eta^+) E_0}\eta^- \eta^+ .\label{alpha_rot}
   \end{matrix}
\end{equation}Notice, equations (\ref{alpha_rot}) provide 12 polarizability components of a/an particle/array when located between two different media. The remaining components can be calculated in a similar way using $E_y'^\pm=E_0 e^{j(\omega t\mp k^\pm x')}$ and $E_z'^\pm=E_0 e^{j(\omega t\mp k^\mp y')}$ as the excitation fields (red and brown colors in Figs.~\ref{test_setup}-b or~\ref{test_setup}-c). The next step is to convert these polrizability components from $x'-y'-z'$ to $x-y-z$ coordinate system. Indeed, the rotation matrix which defines the relation between the rotated and original coordinate systems may be found as~\cite{Arfken}
\begin{equation}\label{RotMat}
    \begin{bmatrix} x' \\[0.3em] y'\\[0.3em] z' \end{bmatrix}
 =R(\theta , \phi)\cdot \begin{bmatrix}
      x \\[0.3em] y \\[0.3em] z \end{bmatrix},
\end{equation} where
$$R(\theta , \phi) = \begin{bmatrix}
       cos \phi 		& 0 		& sin \phi \\[0.3em]
       0 				& 1       & 0 \\[0.3em]
       -sin \phi       & 0 	&  cos \phi
     \end{bmatrix}\cdot\begin{bmatrix}
       1		& 0 		& 0 \\[0.3em]
       0 				& cos \theta         	  & -sin \theta \\[0.3em]
       0      & sin \theta  	& cos \theta
     \end{bmatrix},
$$and, $\theta$ and $\phi$ are the rotation angles around the $x$- and $y$- axes, respectively. Finally, the polarizability tensors in the rotated coordinate system ($\bar{\bar{\alpha'}}$) can be transformed to the polarizability tensors in the original coordinate ($\bar{\bar{\alpha}}$) using the rotation matrix $R$ by the following transformation (see in Appendix-\ref{App_rot}):\begin{equation}
\bar{\bar{\alpha}}=R^{-1}\cdot \bar{\bar{\alpha'}} \cdot R.
\end{equation}Where $R^{-1}$ is the inverse of the rotation matrix $R$.

The next point is associated with the polarizability tensors of a particle in an array. Since we take the advantage of the local currents in our approach, then we inherently take the interaction effect of other particles in the array into account. Consequently, when we use our approach for a particle in an array, we directly calculate the ``effective'' (or collective) polarizability tensors~\cite{Simovski_interaction} of the array. However, in order to compare our method with other approaches, we present two alternative methods for calculation of the effective polarizability tensors. The first one is the calculation of effective polarizability tensors using the individual polarizability tensors and the interaction constants. The second method is the retrieval of the effective polarizabilty tensors from the reflection and transmission coefficients of the array. These two methods are described in the following section while an example which shows the comparison between these approaches is given in section \ref{Eff_chiral}.
\subsection{Effective Polarizabilty Tensors of Planar Periodic Arrays}\label{theory_section_2}
\subsubsection{Effective Polarizabilities from the Individual Polarizabilities and the Interaction Constants}\label{Eff_int}
In order to analyse a scatterer in the close proximity of other scatterers, the effects of other scatterers may be replaced by the interaction constants. According to equation (\ref{defpm}), the dipole moments are expressed in terms of the local electric and magnetic fields $\textbf{E}^\textrm{loc}$ and $\textbf{H}^\textrm{loc}$. These are the fields acting on the particle. Therefore, in a periodic array, these local fields are the contributions of the incident fields $\textbf{E}^\textrm{inc}$ and $\textbf{H}^\textrm{inc}$ as well as the fields created by other dipoles. We may formalize these contributions as~\cite{Tretyakov1}
\begin{equation}
\begin{array}{ll}
\textbf{E}^\textrm{loc}= \textbf{E}^\textrm{inc} + \bar{\bar{\beta}}^\textrm{ee} \cdot \textbf{p},\\
\\
\textbf{H}^\textrm{loc}= \textbf{H}^\textrm{inc} + \bar{\bar{\beta}}^\textrm{mm} \cdot \textbf{m},\\
\\\end{array}
\label{EH_loc}
\end{equation}
where $\bar{\bar{\beta}}^\textrm{ee}$ and $\bar{\bar{\beta}}^\textrm{mm}$ are the electric and magnetic interaction constants. For a periodic array, they may be expressed as~\cite{Tretyakov1}:
\begin{equation}
\begin{array}{ll}
 \bar{\bar{\beta}}^\textrm{ee} = \begin{bmatrix}
      \beta_{\parallel}&0&0\\
       0&\beta_{\parallel}&0\\
      0&0&\beta_{\perp}\\
    \end{bmatrix},\\
       \\
       \bar{\bar{\beta}}^\textrm{mm} = \frac{\bar{\bar{\beta}}^\textrm{ee}}{\eta^2},\\
       \\
       \beta_{\parallel} \approx -j\frac{\omega}{a^2}\frac{\eta}{4}(1-\frac{1}{jkR_0})e^{-jkR_0},\\
       \\
       \beta_{\perp} \approx -j\frac{\omega}{a^2}\frac{\eta}{2}(1+\frac{1}{jkR_0})e^{-jkR_0}.
       \\\end{array}
       \label{ic}
\end{equation}
Here $a$ is the period of the array and $R_0 =a/1.438$~\cite{Tretyakov1}. Replacing (\ref{EH_loc}) in (\ref{defpm}), results in the dipole moments in terms of the incident fields; i.e.,
\begin{equation}
\begin{bmatrix}
      \textbf{p}\\
      \textbf{m}\\
    \end{bmatrix}=
      \hat{\bar{\bar{\alpha}}} \cdot
   \begin{bmatrix}
      \textbf{E}^\textrm{inc}\\
      \textbf{H}^\textrm{inc}\\
    \end{bmatrix}.
\label{pm_eff}
\end{equation}
Where $\hat{\bar{\bar{{\alpha}}}}$ is the ``effective'' polarizabilty tensor and can be calculated as
\begin{equation}
\hat{\bar{\bar{\alpha}}} =
{\begin{bmatrix}
    \bar{\bar{I}} -   \bar{\bar{\alpha}}^\textrm{ee} \cdot \bar{\bar{\beta}}^\textrm{ee}& -  \bar{\bar{\alpha}}^\textrm{em}  \cdot \bar{\bar{\beta}}^\textrm{mm}\\
      - \bar{\bar{\alpha}}^\textrm{me} \cdot \bar{\bar{\beta}}^\textrm{ee}   & \bar{\bar{I}} -  \bar{\bar{\alpha}}^\textrm{mm} \cdot \bar{\bar{\beta}}^\textrm{mm} \\
    \end{bmatrix}}^{-1} \cdot \begin{bmatrix}
      \bar{\bar{\alpha}}^\textrm{ee} & \bar{\bar{\alpha}}^\textrm{em}\\
     \bar{\bar{\alpha}}^\textrm{me} & \bar{\bar{\alpha}}^\textrm{mm}\\
   \end{bmatrix}.
\label{effective_ic}
\end{equation}
In summary, we first calculate the polarizability tensors of an ``individual'' particle and then converts them to the ``effective'' polarizability tensors through (\ref{effective_ic}) using the interaction constants presented in (\ref{ic}). Note that the aforementioned analysis is true for those periodic arrays which can be modeled by dipole moments only. Moreover, it is assumed that the electric and magnetic dipole moments of all particles are in one plane and orthogonal so that the effects of the electric/magnetic dipoles on the magnetic/electric dipoles are zero; that is, $\bar{\bar{\beta}}^\textrm{em} = \bar{\bar{\beta}}^\textrm{me} = 0$.

\subsubsection{Effective Polarizabilities from the Reflection and Transmission Coefficients}\label{Eff_RT}
In this approach, we retrieve the effective polarizability tensors from the reflection and transmission coefficients of the array. For our purpose, it is enough to use the normal illumination. However, the method can be similarly generalized to the oblique illumination. Let us first consider the array in the $x-y$ plane as in Fig.~\ref{test_setup}-c. Then, we illuminate the array with two electromagnetic plane waves with $x$-polarized electric fields from two opposite directions $-z$ and $+z$. The reflected and transmitted electric fields can be expressed in terms of the dipole moments as~\cite{Felsen}
\begin{equation}
\begin{array}{ll}
\textbf{E}^{\textrm{ref}, \pm} = -j\frac{\omega}{2 a^2} (\eta \textbf{p} \pm \hat{\textbf{z}} \times \textbf{m}),\\
\\
\textbf{E}^{\textrm{trans}, \pm} = \textbf{E}^{\textrm{inc},\pm}-j\frac{\omega}{2 a^2} (\eta \textbf{p} \mp \hat{\textbf{z}} \times \textbf{m}),\\
\end{array}
\label{RT}
\end{equation}
where we distinguish between the $\pm z$ illumination direction by the $\pm$ sign. Substituting (\ref{pm_eff}) into (\ref{RT}) results in
$$
\begin{array}{ll}
\textbf{E}^{\textrm{ref},\pm}_{x} = -j\frac{\omega}{2 a^2} (\eta p_x \mp m_y)=\\ \\ -j\frac{\omega}{2 a^2}(\eta \hat{\alpha}_{xx}^\textrm{ee} E_x^\textrm{inc} + \eta \hat{\alpha}_{xy}^\textrm{em} H_y^\textrm{inc} \mp \hat{\alpha}_{yx}^\textrm{me} E_x^\textrm{inc} \mp \hat{\alpha}_{yy}^\textrm{mm}H_y^\textrm{inc})= \\ \\
-j\frac{\omega}{2 a^2}(\eta \hat{\alpha}_{xx}^\textrm{ee} \pm \hat{\alpha}_{xy}^\textrm{em} \mp \hat{\alpha}_{yx}^\textrm{me} - \frac{1}{\eta }\hat{\alpha}_{yy}^\textrm{mm})E_x^\textrm{inc},
\end{array}
$$
for the $x$-component of the reflected electric field. Therefore, the co-component of the reflection coefficient reads as
\begin{equation}
R_\textrm{co}^{\pm}= \frac{E_x^\textrm{ref}}{E_x^\textrm{inc}}=-j\frac{\omega}{2 a^2}(\eta \hat{\alpha}_{xx}^\textrm{ee} \pm \hat{\alpha}_{xy}^\textrm{em} \mp \hat{\alpha}_{yx}^\textrm{me} - \frac{1}{\eta }\hat{\alpha}_{yy}^\textrm{mm}).
\label{Rco}
\end{equation}
Similar formulas can be derived for the co-component of the transmission coefficient and cross-component of the reflection and transmission coefficients as\begin{equation}
\begin{array}{ll}
T_\textrm{co}^{\pm}= \frac{E_x^\textrm{trans}}{E_x^\textrm{inc}}=1-j\frac{\omega}{2 a^2}(\eta \hat{\alpha}_{xx}^\textrm{ee} \pm \hat{\alpha}_{xy}^\textrm{em} \pm \hat{\alpha}_{yx}^\textrm{me} + \frac{1}{\eta }\hat{\alpha}_{yy}^\textrm{mm}),\\ \\
R_\textrm{cr}^{\pm}=\frac{E_y^\textrm{ref}}{E_x^\textrm{inc}}= -j\frac{\omega}{2 a^2}(\eta \hat{\alpha}_{yx}^\textrm{ee} \pm \hat{\alpha}_{yy}^\textrm{em} \pm \hat{\alpha}_{xx}^\textrm{me} + \frac{1}{\eta }\hat{\alpha}_{xy}^\textrm{mm}),\\ \\
T_\textrm{cr}^{\pm}= \frac{E_y^\textrm{trans}}{E_x^\textrm{inc}}=-j\frac{\omega}{2 a^2}(\eta \hat{\alpha}_{yx}^\textrm{ee} \pm \hat{\alpha}_{yy}^\textrm{em} \mp \hat{\alpha}_{xx}^\textrm{me} - \frac{1}{\eta }\hat{\alpha}_{xy}^\textrm{mm}).\\ \\
\end{array}
\label{TRcr}
\end{equation}
In the next step we may solve these eight equations in (\ref{Rco}) and (\ref{TRcr}) in order to retrieve the eight unknown ``effective'' polarizabilies. The results read as\begin{equation}
\begin{array}{ll}
\hat{\alpha}_{xx}^\textrm{ee}=j\frac{a^2}{2 \eta \omega}(\Sigma T_\textrm{co}+\Sigma R_\textrm{co}-2),\\  \\
\hat{\alpha}_{yy}^\textrm{mm}=j\frac{\eta a^2}{2 \omega}(\Sigma T_\textrm{co}-\Sigma R_\textrm{co}-2),\\  \\
\hat{\alpha}_{xy}^\textrm{em}=j\frac{a^2}{2 \omega}(\Delta T_\textrm{co}+\Delta R_\textrm{co}),\quad \hat{\alpha}_{yx}^\textrm{me}=j\frac{a^2}{2 \omega}(\Delta T_\textrm{co}-\Delta R_\textrm{co}),\\  \\
\hat{\alpha}_{yx}^\textrm{ee}=j\frac{a^2}{2 \eta \omega}(\Sigma T_\textrm{cr}+\Sigma R_\textrm{cr}),\quad \hat{\alpha}_{xy}^\textrm{mm}=-j\frac{\eta a^2}{2 \omega}(\Sigma T_\textrm{cr}-\Sigma R_\textrm{cr}),\\  \\
\hat{\alpha}_{yy}^\textrm{em}=j\frac{a^2}{2 \omega}(\Delta T_\textrm{cr}+\Delta R_\textrm{cr}),\quad \hat{\alpha}_{xx}^\textrm{me}=-j\frac{a^2}{2 \omega}(\Delta T_\textrm{cr}-\Delta R_\textrm{cr}).\\  \\
\end{array}
\label{eff_ret}
\end{equation}
Where $\Sigma A_\textrm{co/cr}=A^+_\textrm{co/cr} + A^-_\textrm{co/cr}$ and $\Delta A_\textrm{co/cr}=A^+_\textrm{co/cr} - A^-_\textrm{co/cr}$. Other ``effective'' tangential polarizability components can be retrieved using $y$ polarized incident waves. Moreover, the effective normal polarizability components ($\hat{\alpha}_{kl}^\textrm{ij}$; where $k$ and/or $l$ =$z$ and i,j= e and/or m) cannot be found using the normal illumination. In order to find for the normal polarizability components, one needs to extend this approach for the oblique incidence. However, it is not achievable without increasing the problem complexity.

In the next section, we focus on different examples to validate our semi-analytical approach in calculation of the individual or effective polarizabilities. In the beginning, we calculate the individual polarizability tensors of a chiral particle in microwave frequencies. Then we continue to calculate the effective polarizability tensors of a planar periodic array of these particles using the local currents and compare the results with two other presented methods in sections \ref{Eff_int} and \ref{Eff_RT}.

\section{Results and Discussion}
\subsection{A Wire Chiral Particle}\label{chi_ind}
\label{ind_chiral}
\begin{figure}[h!]
\centering
\includegraphics[width=0.5\textwidth]{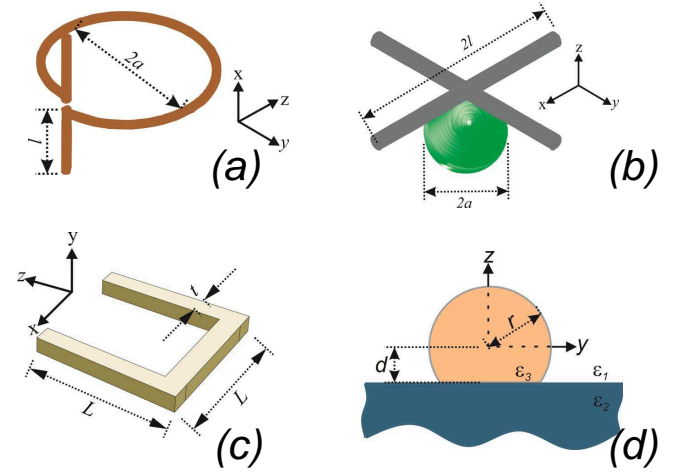}
\caption{The scatterers under study: a) A wire chiral with radius of $a$ and arm length of $l$, the radius of wire is $r_0$, b) A Tellegen-Omega particle composed of two crossed PEC wires with total length of $2l$ on top of a ferrite sphere with radius of $a$, the ferrite sphere is biased along $z$-axis, c) A split ring resonator with an arm length of $L$ and the thickness of $t$, d) A truncated gold sphere on a half space Saffir substrate.}
\label{chiral_geometry}
\end{figure}

\begin{figure*}
\centering
\includegraphics[width=.95\textwidth]{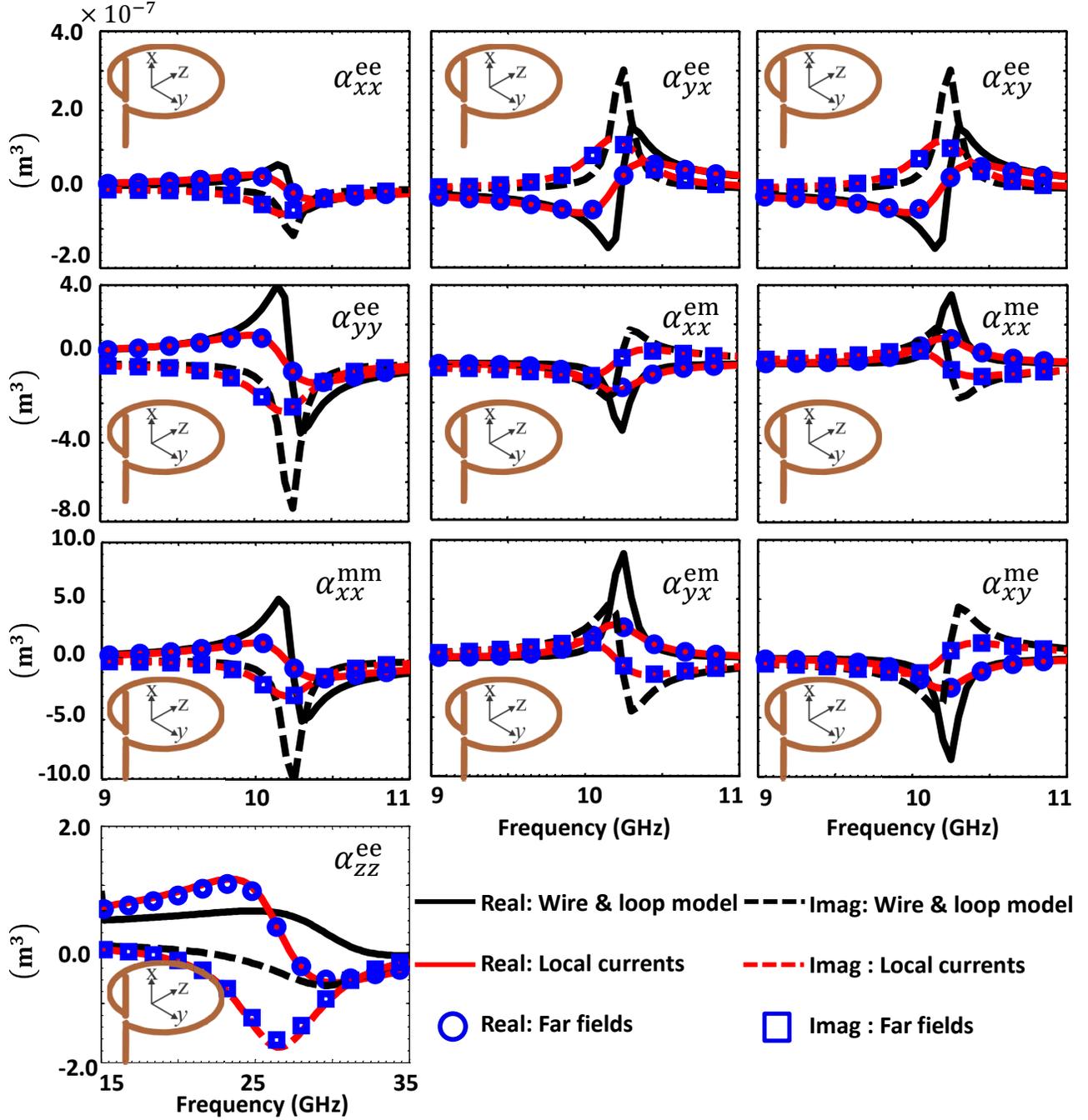}
\caption{The major polarizability components of the chiral element shown in Fig. ~\ref{chiral_geometry} which are calculated by the proposed method and compared to the wire and loop model as well as the far-field method.}
\label{chiral_alpha}
\end{figure*}

One of the interesting bianisotropic particles with a variety of applications in the field of electromagnetics is the ``chiral'' particle~\cite{chiral1, chiral2, chiral3}. A chiral is an object which can be distinguished from its mirror image and cannot be superposed onto it~\cite{Kevin}. A typical wire chiral particle is shown in Fig.~\ref{chiral_geometry}-a. It is composed of a planar loop connected to two straight wires orthogonal to the loop plane.
According to the coordinates of Fig.~\ref{chiral_geometry}-a, this chiral particle provides the following ten major polarizability components\begin{equation}
\begin{array}{ll}
\bar{\bar{\alpha}}^\textrm{ee}\cong \alpha^\textrm{ee}_{xx} \hat{\textbf{x}}\hat{\textbf{x}}+\alpha^\textrm{ee}_{xy} \hat{\textbf{x}}\hat{\textbf{y}}+\alpha^\textrm{ee}_{yx} \hat{\textbf{y}}\hat{\textbf{x}}+\alpha^\textrm{ee}_{yy} \hat{\textbf{y}}\hat{\textbf{y}}+\alpha^\textrm{ee}_{zz} \hat{\textbf{z}}\hat{\textbf{z}},\\\\
\bar{\bar{\alpha}}^\textrm{mm}\cong \alpha^\textrm{mm}_{xx} \hat{\textbf{x}}\hat{\textbf{x}},\\\\
\bar{\bar{\alpha}}^\textrm{em}\cong \alpha^\textrm{em}_{xx} \hat{\textbf{x}}\hat{\textbf{x}}+\alpha^\textrm{em}_{yx} \hat{\textbf{y}}\hat{\textbf{x}},\\\\
\bar{\bar{\alpha}}^\textrm{me}\cong \alpha^\textrm{me}_{xx} \hat{\textbf{x}}\hat{\textbf{x}}+\alpha^\textrm{me}_{xy} \hat{\textbf{x}}\hat{\textbf{y}}.\end{array}
\label{alpha_ch}
\end{equation}
Other polarizability components are negligible or zero compared to these components. Moreover, from the reciprocity we have\begin{equation}
\bar{\bar{\alpha}}^\textrm{ee}=(\bar{\bar{\alpha}}^\textrm{ee})^\textrm{T}, \hspace{0.5cm}\bar{\bar{\alpha}}^\textrm{mm}=(\bar{\bar{\alpha}}^\textrm{mm})^\textrm{T}, \hspace{0.5cm}\bar{\bar{\alpha}}^\textrm{em}=-(\bar{\bar{\alpha}}^\textrm{me})^\textrm{T}.
\label{recip_alpha}
\end{equation}
In Ref.~\cite{Sergei_Chiral}, a wire and loop model is used to analytically calculate the chiral polarizabilities in terms of the model parameters. We choose a wire chiral example with the radius $a = 1.7\textrm{ mm}$ and an arm length of $l = 1.1\textrm{ mm}$ which provides a resonant response around $10\textrm{ GHz}$. The polarizabilities of this particle are calculated using three methods: local currents (our proposed method), the wire and loop model presented in~\cite{Sergei_Chiral}, and the scattered far-field method proposed in ~\cite{Viktar}. The results are depicted in Fig.~\ref{chiral_alpha}.
As can be seen from Fig.~\ref{chiral_alpha}, the results of our method coincide with those of scattered far-field method~\cite{Viktar}. Moreover, the wire and loop model results in an accurate prediction of the resonance frequency while it does not precisely foretell about the strength of the polarizabilities at resonance. The difference is due to the fact that a chiral particle is not exactly a combination of a wire and a loop as considered in~\cite{Sergei_Chiral}. Indeed, they interact together which is neglected in that model. Notice, all results respect the reciprocity theorem through Eq.~\ref{recip_alpha}. Note also the different resonant frequencies of $\alpha^\textrm{ee}_{zz}$ with other polarizability components. This is due to the smaller resonant length of the particle in the $z$ direction.

Next, having these individual polarizabilities in hand, we calculate the effective polarizabilities of a planar periodic array of such chiral elements.

\subsection{A Planar Periodic Array of Chiral Particles}\label{Eff_chiral}
\begin{figure*}
\centering
\includegraphics[width=.95\textwidth]{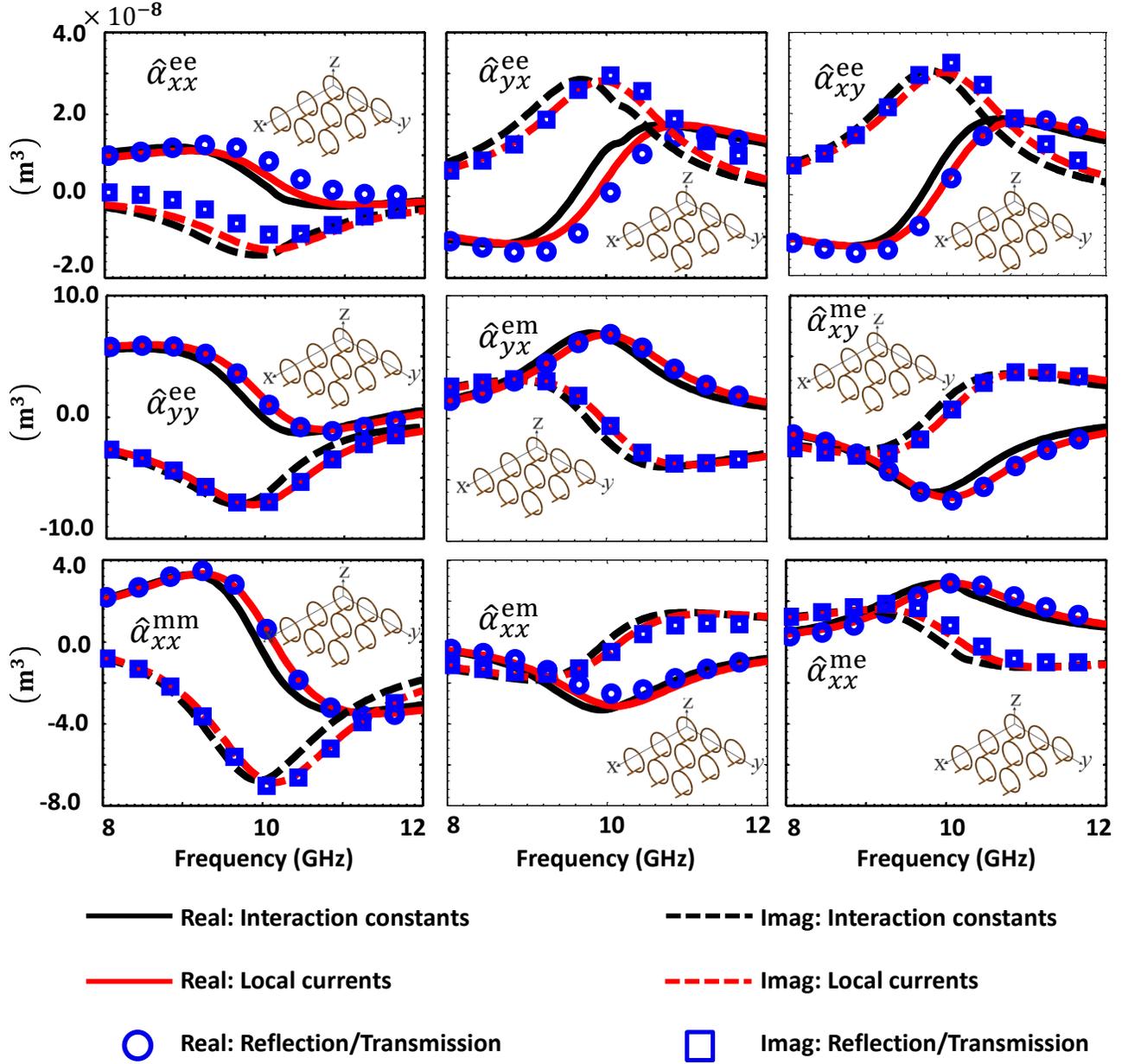}
\caption{The effective polarizabilities of an array of chiral elements with the period of $P=4\textrm{ mm}$. In the proposed method we used integration of local currents over a unit cell of the array while in the interaction constants method we used the individual polarizabilities of Fig.~\ref{chiral_alpha} and calculated the effective polarizabilities using the equation (\ref{effective_ic}). In the retrieving method, we retrieved effective polarizabilities from the reflection and transmission coefficient of the array through equations (\ref{eff_ret}).}
\label{effctive_alpha}
\end{figure*}

We build a planar periodic array composed of similar chiral particles of section \ref{chi_ind} in the $xy$-plane. The period of the array is assumed to be $4\textrm{ mm}$ in both $x$ and $y$ directions. For simplicity and comparison reasons, it is enough to consider only the normal illumination. With this illumination, we may extract all major tangential polarizabilities of the chiral particle given in (\ref{alpha_ch}). However, the normal polarizability $\hat{\alpha}^\textrm{ee}_{zz}$ may not be extracted with normal illumination. Thus, nine major effective polarizabilities from ten important polarizabilities are calculated with three proposed methods: I) our method of direct calculation of the dipole moments through the integration from the local currents over a unit cell of the array, or briefly the method of \textit{local currents}, II) the proposed method which we convert the individual polarizabilities calculated in section-\ref{ind_chiral} to effective polarizabilities using the interaction constants through equation (\ref{effective_ic}), or briefly the method of \textit{interaction constants}, III) the proposed method of reflection and transmission coefficients using equations (\ref{eff_ret}), or briefly the method of \textit{reflection/transmisssion}. The results are plotted in Fig. \ref{effctive_alpha}.
It is obvious from the results, that the three methods are in quite good agreements. However, our first proposed method (\textit{local currents}) is fairly easier to apply since we only need to directly measure the local currents in a unit cell. Moreover, we do not require approximate assumptions (\textit{interaction constants}) or complex calculations (\textit{reflection/transmisssion}). We should mention that the method of \textit{local currents} is independent of the complexity of the problem and we may calculate all $36$ ``effective'' polarizability components without assuming any approximations or complex calculations for the oblique incidence. For instance, in the current example using our first proposed direct method, we may easily calculate the last major polarizability component $\hat{\alpha}^\textrm{ee}_{zz}$. However, the other two methods require either approximations or complex calculations.

Next, we examine the applicability of the method of \textit{local currents} in calculation of the polarizabilities of non-reciprocal scatterers.

\subsection{A Non-reciprocal Tellegen-omega Particle}
\begin{figure*}
\centering
\includegraphics[width=.95\textwidth]{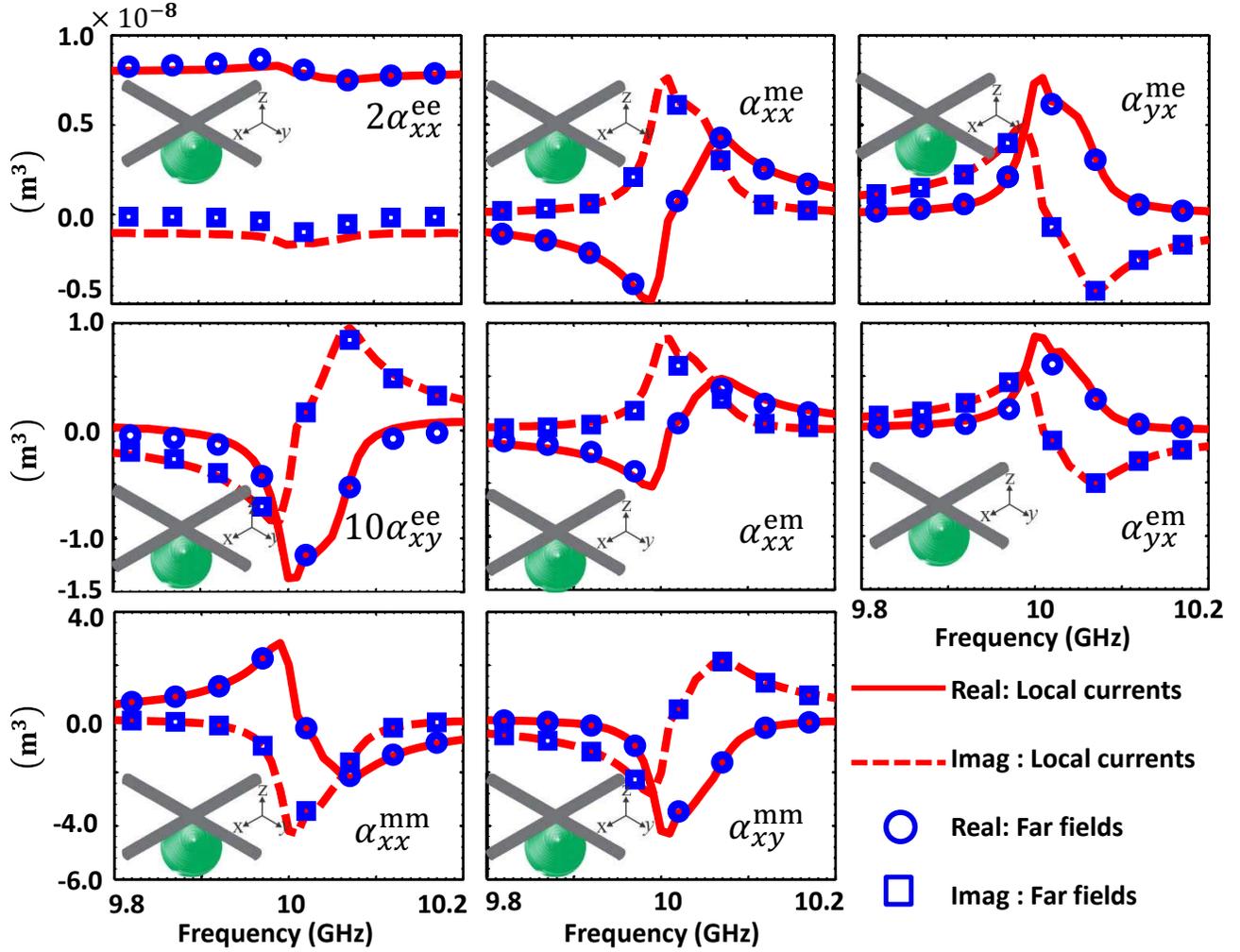}
\caption{Major polarizabilities of the Tellegen-Omega particle of Fig. ~\ref{chiral_geometry}-b calculated through proposed method and compared to far-field approach. The properties and dimensions of the particle are: $l = 1 \textrm{mm}$, $a = 0.5 \textrm{mm}$, Ferrite material: YIG with relative permittivity $\epsilon_r = 15$ and magnetic saturation $M_s = 1780 G$. The applied bias field is $3000 Oe$.}
\label{Tel_OM_alpha}
\end{figure*}

Nonreciprocal particles and their applications are comprehensively discussed in~\cite{Serdyukov}. These particles may be realized using a biased magnetic element (e.g. ferrites) and some conducting wires. Two famous examples are Tellegen-omega/``moving''-chiral particles which simultaneously presents the properties of Tellegen and omega/``moving'' and chiral bianisotropies, respectively. A Tellegen-omega particle as shown in Fig.~\ref{chiral_geometry}-b may be composed of a ferrite sphere and two orthogonal wires. When an incident electric field excites one of the wire arms, the induced electric current will produce two crossed magnetic moments in the ferrite sphere which in turn they induce currents on the wires. From the symmetry of the structure with respect to the $x$ and $y$ axes, one can deduce that\begin{equation}
\begin{array}{ll}
\alpha^\textrm{ee/mm/me/em}_{xx}=\alpha^\textrm{ee/mm/me/em}_{yy},\\
\\
\alpha^\textrm{ee/mm/me/em}_{xy}=\alpha^\textrm{ee/mm/me/em}_{yx}.\\
\\
\end{array}
\label{a_tel}
\end{equation}We calculated the major polarizability components of the structure using the presented method of \textit{local currents} and compared them with the method of scattered far-field presented in~\cite{Viktar}. The results are shown in Fig.~\ref{Tel_OM_alpha}.
Again, for this non-reciprocal element, the results of our proposed method coincide with the scattered far-field results. This proves the power of our method in calculation of the polarizabilities of non-reciprocal particles.

The next example is associated with the calculation of the polarizabilities of an artificial optical magnetic dipole; that is, a nano-gold split-ring resonator (SRR).

\subsection{A Split Ring Resonator}
Split ring resonators (SRRs) are widely used in terahertz frequencies since they provide artificial magnetism which is naturally prohibited at these high frequencies~\cite{Landau}. A typical SRR is depicted in Fig.~\ref{chiral_geometry}-c. Its magnetic moment is created in $y$ direction due to a circulating current in the SRR arms. This current may be induced by an electromagnetic plane wave illumination.

\begin{figure}
\centering
\includegraphics[width=.45\textwidth]{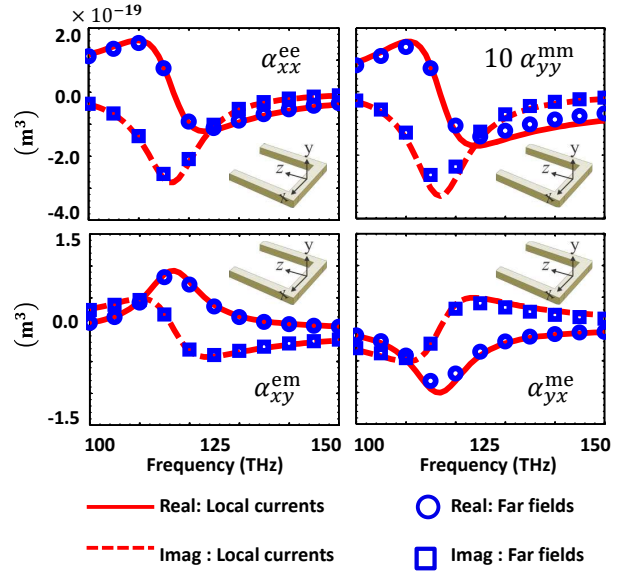}
\caption{The Major polarizabilities of the proposed split ring resonator. Comparison of our method with the far-field method. The length and the thickness of the SRR are $L = 300\textrm{ nm}$ and $t = 40\textrm{ nm}$ while it is made of gold in which its dispersive material parameters are taken from~\cite{Johnson}.}
\label{SRR_alpha}
\end{figure}

The main polarizability components of the proposed SRR are calculated using our method of \textit{local currents} and are compared with the method of scattered far-field presented in~\cite{Viktar}. The results are shown in Fig.~\ref{SRR_alpha}.
As can be seen from the figure, the results of the proposed method coincides with the results extracted from the approach of the scattered far-field. Therefore, one may deduce that the SRR can be replaced with an electric and a magnetic dipole.

\begin{figure}
\centering
\includegraphics[width=.45\textwidth]{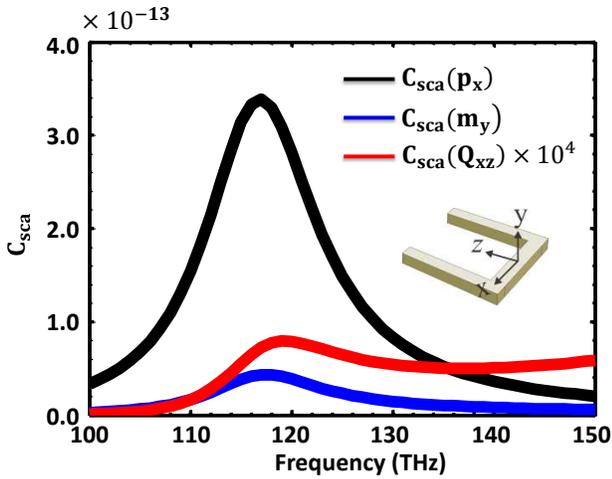}
\caption{Scattering cross sections due to the different multipole moments of the SRR.}
\label{Csca}
\end{figure}

At this step, we clarify an important issue concerning the calculations of higher order multipoles. It is impossible to investigate the effect of higher order multipoles using the method proposed in Ref.~\cite{Viktar} from the scattered far fields. Moreover, to calculate higher order multipoles by the expansion of spherical harmonics using the scattered fields one requires to solve a complicated problem as the order increases (e.g. see in Ref.~\cite{Mulig}). However, since in our proposed method we are numerically integrating from the induced charges/currents over the particle, then we may directly calculate higher order multipoles up to any desired order without increasing the problem complexities. A common way to investigate the effect of higher order multipoles is to compare the effect of different multipole moments in the scattering cross section of the particle~\cite{Mulig}. As the orders of multipoles increase their contributions in the scattering cross section decrease. Here, we compare the effect of strongest electric ($\textrm{p}_\textrm{x}$) and magnetic dipole moments ($\textrm{m}_\textrm{y}$) with the strongest electric quadropole moment ($\textrm{Q}_{\textrm{xz}}$) in Fig.~\ref{Csca}. As can be seen from this figure, the effect of quaropole moments are negligible and therefore SRR can be effectively modelled as a pair of electric and magnetic dipole. This is an important fact that we may investigate directly using our proposed method of local currents. This way we may have a fast physical insight of the complexity of a desired particle. Indeed, without this analysis, it is impossible to consider a particle as dipoles only. As a result, our proposed method not only provides the polarizabilities of any kind of dipole particles but also present a very useful and powerful physical insight of any general particle. More importantly, we do not need extra calculations to grasp such physical insight. Therefore, we gain too many information with a simple method.

Next, we present the potential of our proposed method for particles supported by a refractive substrate.

\subsection{Particles on Substrates}
It is very famous, practical and essential to find the polarizabilities of a substrated particle \cite{Wind, Pederson, Albooyeh_2011}. Indeed, in most practical cases the particles cannot be freestanding and should be supported by a substrate for mechanical robustness. The substrate effect on the polarizabilities of a particle is investigated in many works~\cite{SIB, Albooye_SIB1, Albooye_SIB}. Indeed, it is impossible to apply the proposed scattered far-field method presented in~\cite{Viktar} to calculate for the polarizabilities of a substrated particle. There are some limited analytical works (see e.g. in~\cite{Wind, Pederson}) which solved this problem for a substrated gold truncated sphere. They have solved laplace equations and applied appropriate boundary conditions. Although in their works the electrical polarizabilities in the co and cross directions were calculated, three other polarizabilities (i.e., magnetic, magnetoelectric,and electromagnetic) were neglected. We show here that some of these components are comparable to their electric counterparts. This is due to the presence of the substrate which may lead to a kind of bianisotropy called substrate-induced bianisotropy (SIB)~\cite{Albooyeh_2011}. Therefore, we calculate all polarizabilities of a substrated particle to examine the effect of SIB. We apply the proposed method of local currents to calculate the polarizabilities while we consider the concept of rotation matrix introduced in section\ref{rot_mat} to the structure presented in Refs.~\cite{Wind, Pederson}. This structure is shown in Fig.~\ref{chiral_geometry}-d. It is a truncated gold nano-sphere on a Saffir substrate ($\epsilon_r = 3.13$). We create different geometries by changing the truncating ratio ($ \frac{d}{r}$). The negative and positive ratios are related to the island and cap shapes, respectively. When $ \frac{d}{r} = 0 $, we have a hemisphere and with $ \frac{d}{r} = -1 $, we form a complete sphere. As the structure is symmetric in the $xy$-plane, one can define normal and parallel polarizabilities as $\alpha_{\parallel} = \alpha_{xx} = \alpha_{yy}$ and $\alpha_{\perp} = \alpha_{zz} $. Moreover, all calculated polarizabilities are normalized to the volume of corresponding truncated sphere.

\begin{figure}
\centering
\includegraphics[width=.45\textwidth]{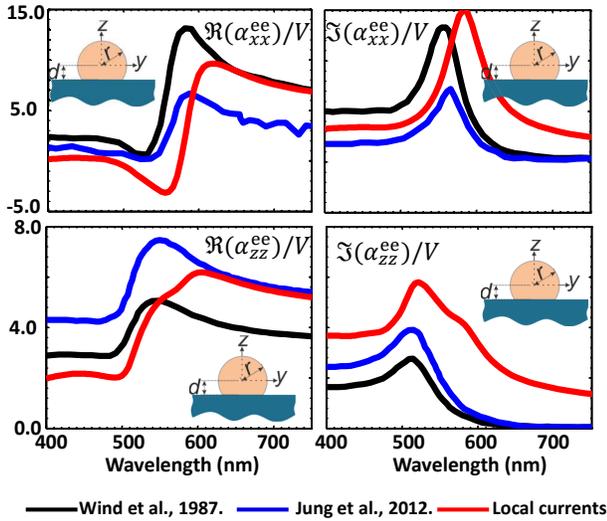}
\caption{The electrical polarizabilities of a gold truncated sphere on a Saffir substrate shown in Fig.~\ref{chiral_geometry}-d with $\frac{d}{r} = -0.4$.}
\label{Sphere_alpha}
\end{figure}

The normalized parallel and normal electrical polarizabilities for $\frac{d}{r}=-0.4$ are plotted in Fig.~\ref{Sphere_alpha} using three different method; that is, our local currents, the method proposed in~\cite{Wind}, and the presented method in~\cite{Pederson}.
The results of these three methods show similar frequency behavior with some differences. The differences between the methods of Refs.~\cite{Wind} and~\cite{Pederson} are due to approximation used in their numerical integrations. {For example, in~\cite{Wind}, it is stated that the results may be deviated up to 13.3 percent with changing the number of multipoles taken into account in particle-substrate interaction.} However, the disagreements between our method and the other two methods is due to the fact that they have neglected the effect of bianisotropy induced by the substrate which we present in the following.

\begin{figure}
\centering
\includegraphics[width=.45\textwidth]{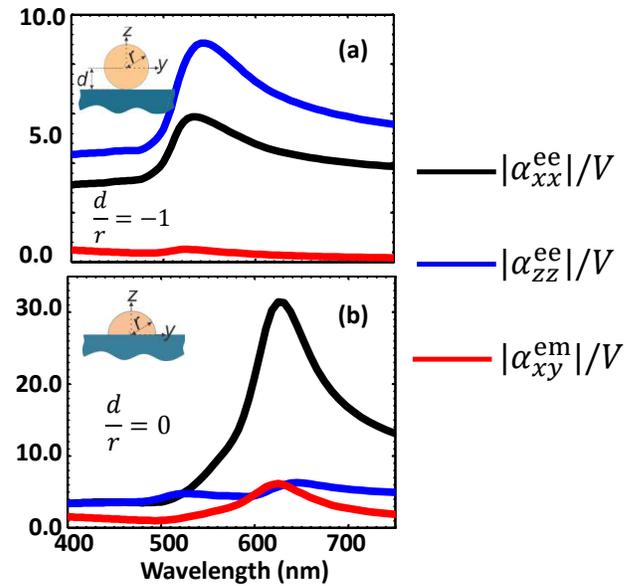}
\caption{a) The magnetoelectric polarizability of a sphere with $\frac{d}{r} = - 1$ compared to its normal and parallel polarizabilities, b) the same figure for a hemisphere $\frac{d}{r} = 0$.}
\label{Sphere_em}
\end{figure}

As we discussed earlier, our proposed method can provide all polarizability components of the structure. With our analysis, we observe that there is a magnetoelectric polarizability component in the $xy$ plane which is comparable to the previously calculated electrical polarizabilities in Refs.~\cite{Wind, Pederson}. Although it is small for small truncation ratio, it starts to be significant as we increase the truncation ratio. For instance, it is about one tenth of the parallel electric polarizability component at the resonance for a complete sphere $\frac{d}{r} = -1$ (Fig.~\ref{Sphere_em}-a) while it increases to one fifth of the parallel electric polarizability component for a hemisphere $\frac{d}{r} = 0$ (Fig.~\ref{Sphere_em}-b).
\begin{figure}
\centering
\includegraphics[width=.45\textwidth]{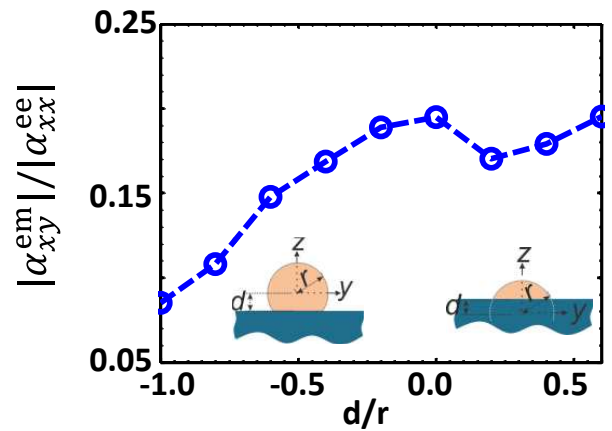}
\caption{The ratio of magnetoelectric polarizability to the parallel electric polarizability for different truncation ratios.}
\label{Sphere_em_rat}
\end{figure}
Notice, the magnetic polarizability (not shown here) is much smaller than the plotted polarizabilities in Fig.~\ref{Sphere_em}-a,b and may be correctly neglected. Figure.~\ref{Sphere_em_rat} demonstrates the effect of the truncation ratio on the magnetoelectric polarizability component. As a result, we prove that the substrate has certainly an inportant effect on the polarizabilitry components of a particle which should be considered in modeling. With this example we conclude our study on the calculation of polarizability components of different type of particles under different circumstances.

\section{Conclusion}
We presented a method for the calculation of polarizability tensors of any arbitrary isolated scatterer, scatterers on substrates and scatterers in arrays. In our method, we used induced local charges/currents of the particle excited by plane waves. The proposed method does not require complicated calculations and harmonic expansions. We expressed different examples (wire chiral, Tellegen-omega, and SRR) and calculated their individual polarizabilities. The results were in quite good agreement with other available methods. We showed how our method present a physical insight of each design and to what extend one may use the concept of very dipole model. Moreover, in the framework of effective polarizabilities of particles in arrays, we compared the presented method with two other distinct methods that we formulated. We finally examined the capability of the proposed method in dealing with substrated particles using an example of truncated sphere on a substrate. Indeed, we investigated that the presence of the substrate lead to a substrate-induced bianisotropy which should not be underestimated.

Generally with our approach, we filled two incomplete parts in the field of calculation of the polarizability tensors; that is, firstly calculation of the polarizabilities for substrated particles and secondly calculation of the effective polarizabilities of a periodic planar array.

\appendices
\section{Transformation of polarizability tensors from one coordinate to a rotated coordinate}
\label{App_rot}
We introduced the rotation matrix in the manuscript through equation (\ref{RotMat}). Therefore, we can transform every vector in the rotated coordinate $x'-y'-z'$ to the original coordinate $x-y-z$ through $\textbf{A}'=R(\theta , \phi)  \textbf{A}$. Consequently, for the fields-moments relation in the rotated coordinate system; i.e.,
\begin{equation}
\begin{bmatrix}
     \textbf{p}'\\
      \textbf{m}'\\
    \end{bmatrix}=\bar{\bar{\alpha'}} \cdot \begin{bmatrix}
      \textbf{E}'\\
      \textbf{H}'\\
    \end{bmatrix},
\end{equation}
one may write:
\begin{equation}
R \cdot \begin{bmatrix}
      \textbf{p}\\
      \textbf{m}\\
    \end{bmatrix}=\bar{\bar{\alpha'}} \cdot R \cdot \begin{bmatrix}
      \textbf{E}\\
      \textbf{H}\\
    \end{bmatrix},
\end{equation}
or equivalently:
\begin{equation}\label{RotOrg}
 \begin{bmatrix}
      \textbf{p}\\
      \textbf{m}\\
    \end{bmatrix}=R^{-1} \cdot \bar{\bar{\alpha'}} \cdot R \cdot \begin{bmatrix}
      \textbf{E}\\
      \textbf{H}\\
    \end{bmatrix},
\end{equation}
Note, vectors without prime sign are original vectors and with prime sign are vectors in the rotated coordinate system. Finally from (\ref{RotOrg}), one may extract:
\begin{equation}
\bar{\bar{\alpha}}=R^{-1}\cdot \bar{\bar{\alpha'}} \cdot R.
\end{equation}
as the relation between the polarizability tensors in the original and rotated coordinate systems.

%

%






\end{document}